\documentstyle[mprocl]{article}

\def\Journal#1#2#3#4{{#1} {\bf #2}, #3 (#4)}

\def\NCA{\em II Nuovo Cimento}

\def\PRA{{\em Phys. Rev.} A}
\def\CJP{\em Can.J.Phys.}
\def\NSC{\em New Scientist}
\def\IJL{\em J.Luminescence}
\def\SCA{\em Scientific American}

\def\be{\begin{equation}}
\def\ee{\end{equation}}
\def\bea{\begin{eqnarray}}
\def\eea{\end{eqnarray}}

\begin{document}

\title{CHAOS AND QUANTUMLIKE MECHANICS IN ATMOSPHERIC FLOWS :
A SUPERSTRING THEORY FOR SUPERGRAVITY}

\author{ A.MARY SELVAM }

\address{Indian Institute of Tropical Meteorology,\\
Dr.Homi Bhabha Marg,Pashan,Pune 411 008,India}


\maketitle\abstracts{
The author has identified quantumlike mechanics in
atmospheric flows with intrinsic nonlocal space-time connections
manifested as the selfsimilar fractal geometry to the global cloud cover
pattern concomitant with inverse power law form for power spectra of
temporal fluctuations. Such long-range spatiotemporal correlations are
generic to dynamical systems in nature and are recently identified as
signatures of selforganized criticality, a field of study belonging to the
newly emerging discipline of nonlinear dynamics and chaos. The author
has presented a universal thory of chaos which postulates that spatial
integration of enclosed small scale fluctuations result in the generation
of a hierarchical scale invariant eddy continuum(network) with ordered
two-way energy flow between the scales. The model concepts lead to the
following results. (1) The eddy energy spectrum follows normal distribution
characteristics,i.e.,the square of the eddy amplitude represents the
probability density,a result which is observed in the subatomic dynamics of
quantum systems. (2) Wave-particle duality is attributed to the bimodal
(formation and dissipation) phenomenological form for manifestation of
energy in the bidirectional energy flow intrinsic to eddy circulations,e.g.,
formation and dissipation respectively of clouds in updrafts and downdrafts
of atmospheric eddies. (3) The nested continuum of eddy flow trajectories
follow Kepler's third law of planetary motion. Therefore,inverse square law
form for centripetal force, representing inertial or gravitational force is
intrinsic to the hierarchical eddy continuum. The above model is analogous
to a superstring model where manifestation of matter is visualised as
vibrational modes in stringlike energy flow patterns.}
  
\section{Introduction}
Atmospheric flows exhibit selfsimilar fluctuations on all scales (space-time)
ranging from climate (kilometers/years) to turbulence (millimeters/seconds)
manifested as the fractal geometry to the global cloud cover pattern
concomitant with inverse power law form for power spectra of temporal
fluctuations. Selfsimilar fluctuations implying long-range correlations
are ubiquitous to dynamical  systems in nature and are identified as
signatures of self-organized criticality \cite{ba} . Standard models in
meteorological theory cannot explain satisfactorily the observed
self-organized criticality in atmospheric flows. Also, mathematical models
for simulation and prediction of atmospheric flows are nonlinear and
computer realizations give unrealistic solutions because of deterministic
chaos, a direct consequence of finite precision round-off error doubling
for each iteration of iterative computations incorporated in long-term
numerical integration schemes used for model solutions. An alternative
non-deterministic cell dynamical system model \cite{al} \cite{ma} \cite{am}
predicts the observed self-organized criticality as a direct consequence of
quantumlike mechanics governing flow dynamics.The model concepts show
that the centripetal acceleration representing the inertial mass of
eddy circulation follows inverse square law form analogous to Newton's
third law for planetary motion.

 \section{Model Concepts}
Atmospheric flows is a representative example of turbulent fluid flows.
The model is based on the concept that spatial integration of enclosed
small scale(turbulent eddy) fluctuations result in organized large
eddy circulations.The eddy energy spectrum therefore follows
statistical normal distribution characteristics according to the
Central Limit Theorem. Therefore,the square of the eddy amplitude,i.e.,
variance represents the statistical normal probability density
distribution.Such a result that the additive amplitudes
of eddies,when squared, represent 
probability densities is observed
in the sub-atomic dynamics of quantum systems such as the electron or photon.
Atmospheric flows therefore follow quantumlike mechanical laws. The
root mean square(r.m.s) circulation speed W of large eddy of radius R is
then given in terms of enclosed small scale eddy circulation speed
w and radius r as

\begin{equation}
W^2 = \frac{2}{\pi}\ \frac{r}{R}\ w^2
\end{equation}

The square of the eddy amplitude $W^2$ represents the kinetic energy E
given as(from Eq.1)

\begin{equation}
E=H\nu
\end{equation}

where $\nu_R$(proportional to 1/R ) is the frequency of the large eddy
and H is a constant equal to $\frac{2}{\pi}\!rw^2$ for growth of large
eddies sustained by constant energy input proportional to $w^2$  from
fixed primary small scale eddy fluctuations. Energy content of eddies
is therefore similar to quantum systems which can possess only discrete
quanta or packets of energy content $h\nu$ where h is a universal constant
of nature (Planck's constant) and $\nu$ is the frequency in cycles per
second of the electromagnetic radiation.
The relative phase angle between large and turbulent eddies is equal
to r/R and is directly proportional to $W^2$(Eq.1). The phase angle therefore
represents variance and also there is progressive increase in phase
with increase in wavelength. The above relationship between phase angle,
variance and frequency has been identified as Berry's Phase \cite{be} in
the subatomic dynamics of quantum systems.Writing Eq(1) in terms of the
periodicities $T(=2 \pi R / W)$and $ t(=2 \pi r / w)$ of large and
small eddies respectively we obtain

\begin{equation}
\frac{R^3}{T^2}\!=\frac{2}{\pi}\!\frac{r^3}{t^2}\!
\end{equation}

Eq.(3) is analogous to Kepler's third law of planetary motion,
namely,the square of the planet's year(period) to the cube of the mean
distance from the Sun is the same for all planets and results in inverse
square law analogous to the Newton's inverse square law for gravitation
\cite{we} . The model is similar to a superstring model for subatomic
dynamics \cite{ka} which unifies quantum mechanical and classical concepts
and incorporates gravitational forces along with nuclear and
electromagnetic forces.The cumulative sum of centripetal forces in a
hierarchy of vortex circulations may result in the observed inverse
square law form for gravitational attraction between inertial masses
(of the eddies). The apparent paradox of wave-particle duality in
microscopic scale quantum systems is however physically consistent
in the context of macroscale atmospheric flows since the bi-directional
energy flow structure of a complete atmospheric eddy results in the
formation of clouds in updrafts region and dissipation of
clouds in downdraft regions.The commonplace occurrence of clouds in a
row is a manifestation of wave-particle duality in the macroscale quantum
system of atmospheric flows. The above-described analogy of quantum-like
mechanics for atmospheric flows is similar to the concept of a subquantum
level of fluctuations whose space-time organization gives rise to the
observed manifestation of subatomic phenomena,i.e.,quantum systems as
order out of chaos phenomena \cite{gg}. H.E.Puthoff \cite{pu} has also
put forth the concept of "Gravity as a zero-point fluctuation force".
The vacuum zero-point fluctuation(electromagnetic) energy manifested
in the Casimir effect is analogous to the turbulent scale fluctuations
whose spatial integration results in coherent large eddy structures.

\section*{Acknowledgments}

The author is grateful to Dr.A.S.R.Murty for his keen interest and
encouragement during the course of the study.

\end{document}